\documentclass[preprint]{acmart}
\usepackage{tabularx}
\usepackage{subfigure}
\usepackage{multirow}
\usepackage{enumitem}
\usepackage{enumitem}
\usepackage{multirow}
\usepackage{array}
\usepackage{enumitem}
\usepackage{nameref}
\usepackage{adjustbox}
\usepackage{xcolor}
\usepackage{tikz}
\usetikzlibrary{positioning}
\usepackage{graphicx}
\AtBeginDocument{%
	}

\setlist{leftmargin=2mm}

\newlist{researchq}{description}{1}
\setlist[researchq,1]{labelwidth=\widthof{\bfseries RQ\ref{rqi}}, leftmargin=!}

\newcolumntype{P}[1]{>{\raggedright\arraybackslash}p{#1}}
\newcolumntype{M}[1]{>{\raggedright\arraybackslash\ttfamily}m{#1}}
\newcommand{\todo}[1]{}
\renewcommand{\todo}[1]{\pdfliteral{1 0 0 rg}#1\pdfliteral{0 0 0 rg}}
\newcommand{\ns}{Not Stemmed}

\newcommand{\vs}{Vocabulary Stemming }
\newcommand{\cs}{Contextual Stemming }
\newcommand{\ecs}{Entity-based Contextual Stemming }

\makeatletter
\let\orgdescriptionlabel\descriptionlabel
\renewcommand*{\descriptionlabel}[1]{%
	\let\orglabel\label
	\let\label\@gobble
	\phantomsection
	\protected@edef\@currentlabel{#1\unskip}%
	\let\label\orglabel
	\orgdescriptionlabel{#1}%
}
\makeatother

\begin{document}
\title[Large Language Models for Stemming: Promises, Pitfalls and Failures]{Large Language Models for Stemming: Promises, Pitfalls and Failures}

\author{Shuai Wang}
\affiliation{
	\institution{The University of Queensland}
	\streetaddress{}
	\city{Brisbane}
	\state{QLD}
	\country{Australia}}
\email{shuai.wang@uq.edu.au}

\author{Shengyao Zhuang}
\affiliation{
	\institution{CSIRO}
	\streetaddress{}
	\city{Brisbane}
	\state{QLD}
	\country{Australia}}
\email{shengyao.zhuang@csiro.au}

\author{Guido Zuccon}
\affiliation{
	\institution{The University of Queensland}
	\streetaddress{}
	\city{Brisbane}
	\state{QLD}
	\country{Australia}}
\email{g.zuccon@uq.edu.au}

\begin{abstract}
	
%In this paper, we investigate the idea of using large language models (LLMs) to stem words. %While stemming is widely used in keyword-based matching pipelines to normalise text before indexing and query processing, these stems are compute for individual terms and have no reliance of contextual information.  
Text stemming is a natural language processing technique that is used to reduce words to their base form, also known as the root form. 
The use of stemming in IR has been shown to often improve the effectiveness of keyword-matching models such as BM25. However, traditional stemming methods, focusing solely on individual terms, overlook the richness of contextual information.

Recognizing this gap, in this paper, we investigate the promising idea of using large language models (LLMs) to stem words by leveraging its capability of context understanding.
With this respect, we identify three avenues, each characterised by different trade-offs in terms of computational cost, effectiveness and robustness : (1) use LLMs to stem the vocabulary for a collection, i.e., the set of unique words that appear in the collection (vocabulary stemming), (2) use LLMs to stem each document separately (contextual stemming), and (3) use LLMs to extract from each document entities that should not be stemmed, then use vocabulary stemming to stem the rest of the terms (entity-based contextual stemming).
%The contextual stemming technique gives rise to the possibility of deciding whether to stem a word or not, which is useful to distinguish between cases in which the word refers to specific entities such as a brand name, place, person name.
Through a series of empirical experiments, we compare the use of LLMs for stemming with that of traditional lexical stemmers such as Porter and Krovetz for English text.  We find that while vocabulary stemming and contextual stemming fail to achieve higher effectiveness than traditional stemmers, entity-based contextual stemming can achieve a higher effectiveness than using Porter stemmer alone, under specific conditions. 

%Code and results are made available at \url{https://anonymous.4open.science/r/SIGIR-2024-LLM-Stemming-7315}.
%or statistical stemmers for European languages such as Hungarian and Czech. 

%We highlight when LLM-based stemmer succeed and importantly when they fail, as well as the costs associated with using LLMs for stemming.

\end{abstract}

\maketitle

\section{Introduction}

Stemming, the process of reducing a word to its \emph{root} form (e.g., `programmer' and `programs' to the common stem `program'), is a common text pre-processing step in many information retrieval (IR) pipelines~\cite{croft2010search,dietz2017component,silvello2018statistical}. Common stemming approaches include algorithmic stemmers, like the Porter~\cite{porter1980algorithm} and Snowball~\cite{porter2001snowball} stemmers that use rule-based heuristics, and dictionary-based stemmers like Krovetz~\cite{krovetz1993viewing} that rely on a predefined list of word forms and their corresponding roots.
The underlying intuition of stemming is to minimise lexical mismatches between terms in queries and documents, thereby improving retrieval effectiveness, especially for lexical-based retrieval algorithms such as BM25. 
Lexical-based retrieval algorithms still form the basis of many more complex IR pipelines that might involve learned rankers, and heavily influence the recall of documents. Stemming is a non-trivial problem, complicated by linguistic nuances, morphological diversity, and context sensitivity.
However, conventional stemming approaches primarily focus on individual terms in isolation, often overlooking the crucial aspect of contextual dependency that can significantly influence the meaning and appropriate stemming of words~\cite{peng2007context}. 

This paper explores the utility of using instruction-based large language models (LLMs) for the task of text stemming within an IR pipeline. LLMs such as  ChatGPT\footnote{\url{https://openai.com/products/chatgpt}.}, LlaMa~\cite{touvron2023llama}, Falcon~\cite{penedo2023refinedweb}, SOLAR~\cite{kim2023solar} have shown great potential across several natural language processing tasks including question answering~\cite{yang2023large,kopf2023openassistant,singhal2023towards,chang2023survey}, query processing~\cite{ws2023chatgpt,wang2023generating} and document summarization~\cite{yang2023exploring,lin2023adapting}, and promise across several IR tasks such as query generation and augmentation~\cite{inpars,inparsv2,dai2023promptagator,zhuang2023augmenting}, ranking~\cite{gao2022precise,ma2023zero,pradeep2023rankvicuna,qin2023large,sachan-etal-2022-improving,sun2023chatgpt,zhuang2023setwise,wang2024zero} and evaluation~\cite{faggioli2023perspectives,thomas2023large}. In using LLMs for stemming, we initially identify two avenues:  \textit{Vocabulary Stemming} (VS),  where we use LLMs to stem the whole vocabulary of the collection (and the query), and \textit{Contextual Stemming} (CS), where we use LLMs to stem each document (or query) independently of other documents. CS allows for stemming to be contextual to that document, i.e. the same word appearing in different documents might be stemmed in one but not in the other (or stemmed to a different root). We then provide a further refinement to the CS method, by introducing an \textit{Entity-based Contextual Stemming} (ECS), in which the LLM is forced to recognise entities and only apply stemming to non-entity words.

To study the effectiveness of using LLMs for stemming, we provide initial results on experimenting with three state-of-the-art LLMs: \texttt{GPT-3.5-turbo-0613} (ChatGPT), \texttt{LlaMa-2-7b-chat}~\cite{touvron2023llama2} and \texttt{SOLAR-10.7B-instruct}~\cite{kim2023solar}. Our results show the promise and failures of these LLM-based stemmers, providing insights on their effectiveness.

\newcommand{\rqi}{How does the effectiveness of LLM-based stemming methods compare to traditional stemming methods?}
\newcommand{\rqii}{Does \textit{Contextualised Stemming} yield with a higher effectiveness than \textit{Vocabulary-Based Stemming?}}
\newcommand{\rqiii}{Can \textit{Entity-Based Stemming} outperform \textit{Contextualised Stemming} by simplifying LLM task complexity?}
\newcommand{\rqiv}{What is the consistency of performance gains or losses when comparing LLM-based stemmers to traditional stemmers?}

%\begin{description}
%	\item[RQ1\label{rqi}] \rqi
%	
%	\item[RQ2\label{rqii}] \rqii
%
%	
%	\item[RQ3\label{rqiii}] \rqiii
%	
%	\item[RQ4\label{rqiv}] \rqiv
%	
%
%\end{description}

%\todo{following still waiting for result to come out}

%\input{related_work}
\section{Stemming Using LLMs}

\begin{figure*}[t!]
	\centering

	\includegraphics[width=0.9\textwidth]{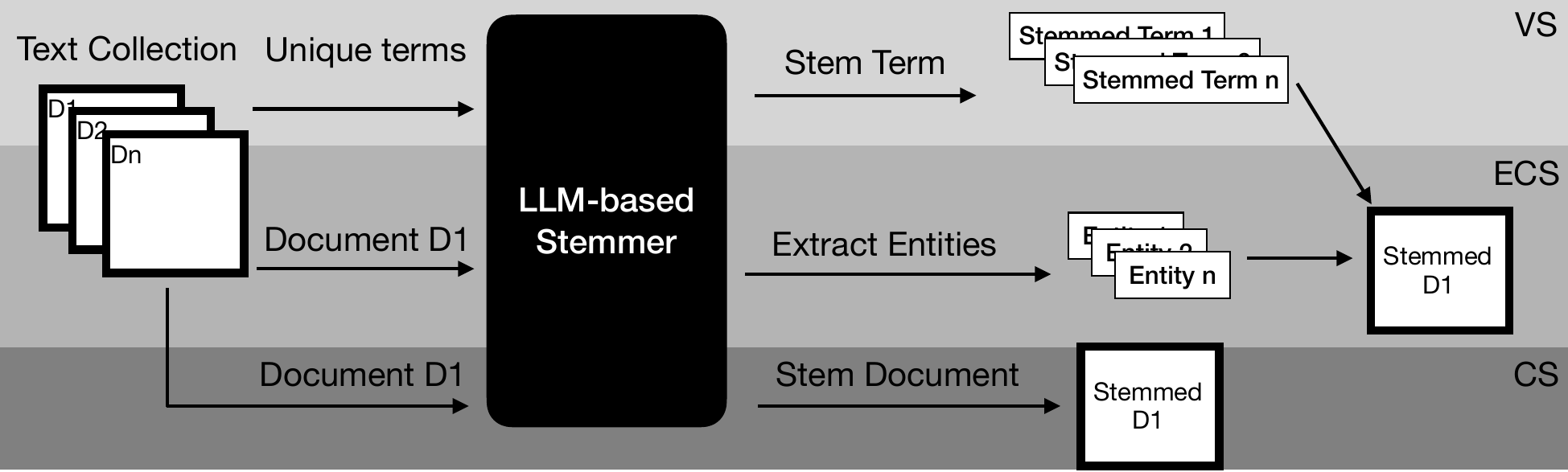}

	\caption{Pipeline for stemming techniques described in the study.}

	%$ P(\texttt{no}|d,t)$ is the probability of the \texttt{no} token,
	\label{fig:architecture}
\end{figure*}

We investigate using LLMs for stemming by examining three approaches, visualized in Figure~\ref{fig:architecture}. 

 \vs (VS) operates under the assumption that each term in a document or query is stemmed individually, irrespective of the context in which it appears, similarly to traditional stemming approaches like Porter or Krovetz stemmers. We thus leverage LLMs to obtain the root form for each word. For efficiency reasons, we first tokenize all documents (and queries) into words, then create a lexicon containing each unique word (the vocabulary), and then we submit each word in the vocabulary to the LLM via the corresponding prompt in Table~\ref{table:prompt} to obtain its stemmed version.

%In our work, we investigate how effective can large language model help to stem the queries and collections, so that it can help with  a better ranking with respect to lexical-based ranking method, such as BM25~\cite{robertson2009probabilistic}.
%We introduce three variations for stemming through LLMs: \textit{Vocabulary Stemming}, \textit{Contextual Stemming} and \textit{Entity-Based Contextual Stemming}. We show the prompt used for the stemming methods in Table~\ref{table:prompt}.
%
%\subsection{\vs}
%\vs operates under the assumption that each term in a document or query is stemmed individually, irrespective of its contextual position, simlar to traditional stemming approaches like Porter Stemmer or Krovetz Stemmer. We leverage LLMs to capture the root form for each term. For efficiency reasons, we ground all unique terms in the collection/query into chunks, then prompt the model to stem the terms chunk by chunk.

%\subsection{\cs}

\cs (CS) operates under the assumption that the context in which a word appears should dictate whether stemming should occur, or not.  In this method, the document (or query) is directly fed into the LLM, and the model outputs the stemmed version of the input text. In addition, we instruct the LLM to not stem words associated to people's names, places, organisations, brands etc. -- i.e. entities. 
The advantage that a contextual stemmer might provide is that for the sentence “Programs PTY. LTD. sold for 1 billion euros”, it might recognize “Programs” to refer to an entity (a company in this case): not stemming this word might be a good choice as it would provide a more nuanced and accurate term representation which in turns might improve retrieval effectiveness.
We note that the LLM might produce different stems to the same word appearing in different contexts: while linguistically incorrect, this may allow for a more focused matching.

\ecs (ECS) builds upon the contextual stemming method, but it explicitly focuses on entities. 
Documents (or queries) are still processed independently, but instead of asking the LLM to stem, we instruct the LLM to identify entities. We then use a vocabulary stemmer (either based on LLM or a traditional linguistic stemmer such as Porter) to stem every word but words identified by the LLM as belonging to entities\footnote{Although entities consisting of multiple words are split.}; the intuition being that entities (people's names, places, organisations, brands etc.) should not be stemmed as they would be confused to unrelated occurrences of words with the same stem. This is similar to the CS, but in ECS we only instruct the LLM to identify entities, and leave the stemming to the linguistic tools. We devise two variations of the ECS pipeline. In ECS.1, for words that have been identified as referring to entities, we only include them in their original (not stemmed) form for indexing. In ECS.2, along with including the original (not stemmed) form, we also include their stemmed form. This variation was devised after observing and analysing results for ECS.1 and identifying that certain queries would have benefit from matching  against the stemmed version of entity words.

Note that, as shown by the prompts in Table~\ref{table:prompt}, all these approaches are `one-shot', in that a single example is provided to the LLM to explain the task.

%aims to create a simpler task than Contextualised Stemming, as the sole task for the model is to identify the terms that \textit{should not} be stemmed. The intuition of Entity-Based Stemming is that terms that are named entities, such as names, places, organisations, or brand should not be stemmed, as they often serve as an important indicator about this text, and stemming to its root term could change the semantics of the term and lead to a less effective ranking.

%aims to exploit the underlying semantics of context using LLM, so that the model can decide when should a term being stemmed/not stemmed, depending on the context of a word. In this method, query/document in the dataset will be directly feed into LLM, and model outputs are the stemmed version of the input text. By embedding this contextual information, stemming results in a more nuanced and accurate term representation, thereby potentially boosting the effectiveness of ranking algorithms such as BM25.

\begin{table*}[t!]
	\centering
	\small
	\caption{Prompts for three LLM-based stemming methods. Italicised text indicates variables that are replaced with the respective content.\vspace{-7pt}}
	\resizebox{0.9\textwidth}{!}{
\begin{tabular}{c|M{480pt}}
	\toprule
	 & \textnormal{Prompt} \\
	\hline
	
%	\rotatebox[origin=c]{90}{\parbox{1cm}{\centering ChatGPT}} &
%	For each term provided on a separate line, please: \linebreak
%	1. Split the term if it has extraneous characters.\linebreak
%	2. Apply stemming to reduce the words to their base or root form.\linebreak
%	3. Do not stem names, places, organizations, or brands.\linebreak
%	Given terms:\linebreak${terms}$\linebreak
%	For the output, place result from each original term a new line, formatted by `original term:produced stems'. If a single original term produces multiple stems, separate produced stems with a space. \\ \hline
	%\linebreak \#\#\# Input: \hline
	
	\rotatebox[origin=c]{90}{VS} &You are a professional stemmer that is responsible to stem text. Text stemming is a natural language processing technique that is used to reduce words to their base form, also known as the root form. The process of stemming is used to normalize text and make it easier to process. %You should stem every word except from names of people, places, organisations, brands.
	\linebreak
	Your output should strictly follow the format "original word:stem". If a single original word produces multiple stems, separate the stems with a space.\linebreak
	Can you provide the stemmed version of these terms?
	$\{terms\_sample\}$ \linebreak
	Stemmer: $\{stemmed\_sample\}$ \linebreak
	Can you provide the stemmed version of these terms? $\{terms\}$ \linebreak
	Stemmer:
	\\ \hline
	\rotatebox[origin=c]{90}{CS} &You specialize in text stemming, text stemming is a natural language processing technique that is used to reduce words to their base form, also known as the root form. The process of stemming is used to normalize text and make it easier to process. You should stem every word except from names of people, places, organisations, brands. For example, the words "programming,” "programmer,” and "programs” can all be reduced down to the common stem "program”. However, in the sentence "Programs PTY. LTD. sold for 1 billion euros”, the word "programs” should not be stemmed because it refers to the name of a company. \linebreak
	Can you provide the stemmed version of this paragraph?
	$\{paragraph\_sample\}$\linebreak
	Stemmed paragraph:
	$\{stemmed\_paragraph\_sample\}$\linebreak
	Can you provide the stemmed version of this paragraph?
	$\{paragraph\}$\linebreak Stemmed paragraph:
	\\ \hline
	
	\rotatebox[origin=c]{90}{ECS} &You specialize in identifying and preserving entities, such as names, brands, or organisations, within text paragraphs. It's imperative to ensure these terms are not stemmed to enhance search engine performance. \linebreak
	Can you extract all entities from the following paragraph?
	$\{paragraph\_sample\}$\linebreak
	Extracted entities:
	$\{entities\_extracted\_sample\}$\linebreak
	Can you extract all entities from the following paragraph?
	$\{paragraph\}$\linebreak Extracted entities:
	\\
	\bottomrule
\end{tabular}
}
\vspace{-7pt}
	\label{table:prompt}

\end{table*}

%\subsection{\ecs}
%\ecs  aims to create a simpler task than Contextualised Stemming, as the sole task for the model is to identify the terms that \textit{should not} be stemmed. The intuition of Entity-Based Stemming is that terms that are named entities, such as names, places, organisations, or brand should not be stemmed, as they often serve as an important indicator about this text, and stemming to its root term could change the semantics of the term and lead to a less effective ranking.

\section{Experimental Setup}
%This section delineates the experimental design, including the specific LLMs employed, datasets used, and evaluation metrics adopted to assess the performance.

\textbf{Model Specifications.}
We experimented with three LLMs: \texttt{GPT-3.5-turbo-0613} (ChatGPT), \texttt{LlaMa-2-7b-chat} (LlaMa-2) \cite{touvron2023llama2}, \texttt{SOLAR-10.7B-instruct} (SOLAR) \cite{kim2023solar}. ChatGPT is a prominent commercial language model known for its advanced natural language understanding and generation capabilities. In our experiments, we treat ChatGPT as a black-box tool and utilise OpenAI's API to acquire access to the LLM, and set the temperature to 0. LlaMa-2 is an open-source LLM. The model is optimised for dialogue applications and is able to handle user instructions comprehensively. We set the temperature to 0.000001~\footnote{$temperature=0$ is forbidden in the transformer library.}, $do\_sample=True$, $top\_p=0.9$,  $num\_beams=1$. SOLAR is the leading open-source LLM in the openLLM leaderboard\footnote{Benchmark evaluating the capabilities of LLMs on various NLP tasks.}\cite{open-llm-leaderboard} at the time of writing. SOLAR shares tokenizer and chat templates with LlaMa but features unique architectural elements; we configured it identically to LlaMa-2 for consistent comparison.

We used LlaMa-2 and SOLAR across all our experiments. Experimenting with ChatGPT incurs costs related to API usage. We use ChatGPT only when experimenting with VS, as this resulted in a cost of \$20: experimenting with contextual stemmers would have incurred costs in excess of \$1,000.

%
% and is currently the most effective open-source LLM on leaderboard~\cite{beeching2023open}~\footnote{We set $temperature=0.000001, do\_sample=True, top\_p=0.9,  num\_beams=1$}. 

\textbf{Baseline Stemmers.} LLM stemmers are compared against two common stemmers, Porter and Krovetz, as implemented in Lucene. We also consider the baseline condition of not performing any stemming (\ns). Because \ecs uses LLMs to perform entity extraction, we also experiment by replacing LLMs with a more common Roberta-based named-entity recognition model~\cite{conneau2019unsupervised} (note, this is a fine-tuned model).

\textbf{Ranking Pipeline.} The output of the stemming methods was used to index documents and prepare queries using Pyserini~\cite{lin2021pyserini}. For retrieval, we used the BM25 function with parameters set to default in Pyserini.

%Our experiments utilize BM25 as the target lexical-based ranking algorithm against which the effectiveness of different stemming methods is evaluated. 

%We compare our large language model-based stemming techniques against three established baseline stemming methods: \ns, \ps and \ks. 
%\textit{\ns} approach serves as a control where terms in the document collection and queries are used in their original form, without any stemming.
%For \ps and \ks, we use the Lucene English stemmer in the library.

%\textit{\ps} is a widely used stemming algorithm that truncates words to their root form~\cite{}
%\textit{\ks} focuses on converting English words to their root form by utilizing a rich lexicon and syntactic rules~\cite{}.

%\subsection{Dataset \& Evaluations}
\textbf{Datasets.} Our experiments were constrained by our computing infrastructure and budget. We estimated that experimenting with large datasets like MS-Marco would require over 10,000 GPU hours of an A100 to run a single \cs or \ecs experiment. We then resorted to using two smaller datasets: \textit{Trec-Covid} and \textit{Trec-Robust04-LAT}. The Trec-Covid dataset comprises of 171,332 biomedical documents and 50 test queries. Trec-Robust04-LAT comprises the subset of the TREC Robust04 collection made up by news-wire articles from the LA Times. The dataset contains a total of 171,332 documents and 249 test queries.

%We evaluate our approach on two small-scale datasets~\footnote{we are constrained by our computation resource from large-scale dataset like Msmarco, which requires over 10,000 gpu hours using A100 GPUs for running one experimental setting.}: \textit{Trec-Covid} and \textit{Trec-Robust04-LAtimes}, each comprising fewer than 200,000 documents.

%\paragraph{\textbf{Trec-Covid}} dataset serves as a biomedical information retrieval (IR) corpus specifically designed for advancing zero-shot searching in the context of COVID-19. The corpus encompasses 171,332 documents and includes a set of 50 test queries. On average, each query is associated with approximately 493.5 relevant documents.
%
%\paragraph{\textbf{Robust04-LAT}} dataset is a subset of the original Trec-Robust04 collection and includes only source documents from the LA Times. It is intended for news search tasks. This dataset contains 131,896 documents and 249 topic queries, with an average of roughly 70 relevant documents per topic query.

\textbf{Data Preprocessing.}
%For both datasets, we rely on Pyserini~\cite{lin2021pyserini} pre-built index to obtain raw documents. 
%Due to computational constraints, 
We truncated each document to its first 300 words; we used the truncated text to represent the document. This is akin to using the ``FirstP'' method devised for ranking long documents using BERT-based rankers~\cite{dai2019firstp}. 
This pre-processing step is consistent across all stemmers to minimize bias, and it is introduced to ensure a manageable computational load.

\textbf{Evaluation Measures.} We use Reciprocal rank (RR), MAP, nDCG@10 and Recall@1000, which are common metrics in IR. These metrics offer complementary perspectives on retrieval quality and were used in previous work to evaluate ranking effectiveness for both datasets.

\section{Results}
\label{sec:result}

\begin{table*}[t!]
	\centering
	\small
	\caption{Main result table for LLM-based stemming methods. Statistical significance is assessed using a Student's two-tailed paired t-test with a Bonferroni correction ($p < 0.05$) with respect to the Porter Stemmer, denoted by *\vspace{-5pt}.}
	\begin{tabular}{l|l|l|llll|llll}
	\toprule
	\multirow{2.5}{*}{}&\multirow{2}{*}{\parbox{1.6cm}{Entity\\Extraction\\Model}}&\multirow{2.5}{*}{Stemmer} & \multicolumn{4}{c}{Trec-Covid} & \multicolumn{4}{c}{Trec-Robust04-LAT} \\
	\cmidrule(lr){4-7} \cmidrule(lr){8-11}
	&&& {RR}& {MAP} & {ndcg10} & {R1000} & {RR} & {MAP} & {ndcg10} & {R1000} \\
\midrule

\multirow{3}{*}{Baseline}&N/A&N/A & 0.4297* & 0.0753* & 0.2360* & 0.2096* & 0.2786* & 0.0439* & 0.1261* & 0.1649* \\
&N/A&Krovetz & 0.4553* & 0.0894* & 0.2765* & 0.2347* & 0.3222* & 0.0538* & 0.1536* & 0.1857* \\
&N/A&Porter & 0.8666 & 0.1896 & 0.6018 & 0.3958 & 0.4975 & 0.0853 & 0.2440 & 0.2463 \\ \midrule
\multirow{3}{*}{VS}&N/A&ChatGPT  & 0.7368 & 0.1599* & 0.4975* & 0.3564* & 0.4305* & 0.0746* & 0.2070* & 0.2278* \\
&N/A&LlaMa-2 & 0.4987* & 0.1093* & 0.3439* & 0.2762* & 0.2919* & 0.0454* & 0.1341* & 0.1772* \\
&N/A&SOLAR & 0.6428* & 0.1424* & 0.4319* & 0.3379* & 0.4088* & 0.0656* & 0.1938* & 0.2163*  \\ \midrule
\multirow{2}{*}{CS}&N/A&LlaMa2& 0.3375* & 0.0496* & 0.1736* & 0.1695* & 0.2265* & 0.0335* & 0.1016* & 0.1472* \\
&N/A&SOLAR & 0.6507* & 0.0767* & 0.3905* & 0.2244* & 0.2840* & 0.0358* & 0.1172* & 0.1440* \\ \midrule
\multirow{12}{*}{ECS.1}&Roberta&Porter & 0.8130 & 0.1823 & 0.5658 & 0.3837 & 0.5027 & 0.0822 & 0.2420 & 0.2459 \\
&Roberta&ChatGPT & 0.6383* & 0.1387* & 0.4273* & 0.3278* & 0.4035* & 0.0670* & 0.1872* & 0.2162* \\
&Roberta&LlaMa-2 & 0.4205* & 0.0738* & 0.2279* & 0.2247* & 0.3072* & 0.0464* & 0.1332* & 0.1773* \\
&Roberta&SOLAR & 0.6389* & 0.1289* & 0.4366* & 0.3125* & 0.3621* & 0.0589* & 0.1696* & 0.2008* \\  \cmidrule{2-11}
&LlaMa-2&Porter & 0.8179 & 0.1868 & 0.5719 & 0.3939 & 0.5191 & 0.0879 & 0.2550 & 0.2530 \\
&LlaMa-2&ChatGPT & 0.6851* & 0.1160* & 0.4197* & 0.2925* & 0.4062* & 0.0669* & 0.1941* & 0.2153* \\
&LlaMa-2&LlaMa-2 & 0.3544* & 0.0734* & 0.2066* & 0.2081* & 0.3072* & 0.0465* & 0.1393* & 0.1760* \\
&LlaMa-2&SOLAR & 0.4949* & 0.0894* & 0.2871* & 0.2507* & 0.3529* & 0.0554* & 0.1615* & 0.1886* \\  \cmidrule{2-11}
&SOLAR&Porter & 0.7851 & 0.1808 & 0.5634 & 0.3851 & 0.5059 & 0.0870 & 0.2501 & 0.2472 \\
&SOLAR&ChatGPT & 0.6921 & 0.1218* & 0.4464* & 0.3019* & 0.4030* & 0.0661* & 0.1897* & 0.2141* \\
&SOLAR&LlaMa-2 & 0.4578* & 0.0630* & 0.2452* & 0.2023* & 0.2983* & 0.0421* & 0.1275* & 0.1711* \\
&SOLAR&SOLAR & 0.6382* & 0.0993* & 0.4124* & 0.2683* & 0.3557* & 0.0541* & 0.1658* & 0.1911* \\ \midrule
\multirow{12}{*}{ECS.2}&Roberta&Porter& 0.8134 & 0.1818 & 0.5694 & 0.3766 & 0.4920 & 0.0834 & 0.2423 & 0.2463 \\
&Roberta&ChatGPT & 0.6746* & 0.1590* & 0.4710* & 0.3473* & 0.4270* & 0.0724* & 0.1973* & 0.2264* \\
&Roberta&LlaMa-2 & 0.4242* & 0.0861* & 0.2576* & 0.2413* & 0.3134* & 0.0484* & 0.1371* & 0.1896* \\
&Roberta&SOLAR & 0.6196* & 0.1417* & 0.4042* & 0.3300* & 0.3741* & 0.0641* & 0.1816* & 0.2157* \\  \cmidrule{2-11}
&LlaMa-2&Porter & \textbf{0.9058} & \textbf{0.2410*} & \textbf{0.6623} & \textbf{0.4538*} & \textbf{0.5412} &\textbf{ 0.0962*} & \textbf{0.2727*} & \textbf{0.2614*} \\
&LlaMa-2&ChatGPT & 0.7710 & 0.2178 & 0.5688 & 0.4297 & 0.4249* & 0.0821 & 0.2214 & 0.2384 \\
&LlaMa-2&LlaMa-2 & 0.4425* & 0.1246* & 0.2892* & 0.3002* & 0.3426* & 0.0558* & 0.1659* & 0.1941* \\
&LlaMa-2&SOLAR & 0.6585* & 0.1793 & 0.4669* & 0.3836 & 0.4143* & 0.0714* & 0.2058* & 0.2220* \\ \cmidrule{2-11}
&SOLAR&Porter & 0.7860 & 0.1809 & 0.5642 & 0.3871 & 0.5121 & 0.0921* & 0.2606* & 0.2513 \\
&SOLAR&ChatGPT & 0.6634* & 0.1550* & 0.4466* & 0.3500* & 0.4245* & 0.0778 & 0.2107* & 0.2276* \\
&SOLAR&LlaMa-2 & 0.4182* & 0.0826* & 0.2379* & 0.2398* & 0.3185* & 0.0513* & 0.1421* & 0.1896* \\
&SOLAR&SOLAR & 0.5936* & 0.1375* & 0.4021* & 0.3244* & 0.3817* & 0.0669* & 0.1896* & 0.2133* \\
	\bottomrule
\end{tabular}
	\label{table:result}
\vspace{-5pt}
\end{table*}

%Table~\ref{table:result} shows results obtained in our experiment.

\subsubsection*{LLM-based Stemming VS. Traditional Stemming: }
The results, reported in Table~\ref{table:result}, provide valuable perspectives on the effectiveness of LLM-based stemming methods. While VS generally performed worse than Porter stemmer, it did outperform the Krovetz stemmer across all experiments. Interestingly, despite the intuition that CS might better exploit word context and model text semantics at a document level, and thus make better stemming decisions, it actually performed substantially below expectations, and always less than VS-stemmed when the same LLM is applied.
This could be attributed to the inherent complexity of the task, and show that current LLMs are still not effective in stemming an entire document at once. 
We also highlight that the computational costs associated with CS is a key drawback. Even with a limited dataset and the use of a relatively small model variant, the computational time exceeded 200 hours on one of our test datasets\footnote{Computational cost measured using Nvidia A100 GPU.}, underscoring the impracticality of CS for large-scale document collections.

%We employed two stemming techniques, \vs (VS) and \cs (CS), directly applied to the corpus using LLMs. For the VS method, known for its computational efficiency, we harnessed the OpenAI ChatGPT API in conjunction with the LlaMa-2 model to investigate the influence of different LLMs. In contrast, the CS method, which demands more computational resources, was exclusively paired with the LlaMa-2 model.

%As demonstrated in Table~\ref{table:result}, both the ChatGPT-based and LlaMa-2-based VS stemming approaches outperform the traditional Krovetz stemmer, albeit falling short of the effectiveness exhibited by the Porter stemmer. Moreover, when applied to CS, LlaMa-2's performance decreased significantly, yielding markedly inferior results while demanding considerably more computational power. Thus, we conclude that employing LLM-based stemmers directly, in a manner akin to traditional stemmers, does not yield optimal outcomes.

%\subsection{\vs VS. \cs}

\subsubsection*{Effectiveness of \ecs: }

Our results for ECS provide a nuanced picture. When ECS was applied using the LlaMa-2 model for entity extraction and the Porter stemmer for the remaining terms (referred to as ECS.1), its effectiveness was comparable to that of the standalone Porter stemmer. Specifically, ECS.1 marginally underperformed Porter on Trec-Covid but slightly outperformed it on Trec-Robust04-LAT. However, these differences were not statistically significant. %, indicating that the benefit of using LLMs for entity identification may be dataset-dependent.
Interestingly, when both original and stemmed forms of identified entities were retained (method ECS.2), ECS showed a statistically significant improvement in effectiveness over the Porter stemmer alone with respect to MAP. These findings suggest that incorporating both the stemmed and unstemmed forms of the identified entities is key to create an effective BM25 ranker.

Furthermore, we extended our analysis to compare entity extraction capabilities across LlaMa-2, SOLAR, and Roberta-based models. Despite SOLAR's superior performance in other NLP tasks, it was consistently outperformed by LlaMa-2 in entity extraction. Interestingly, LlaMa-2 also demonstrated a higher performance over the Roberta model in this task, even without task-specific fine-tuning, contrary to the fine-tuned Roberta model.
 %This finding suggests that by designing carefully, LLM-based methods could surpass task-specific models.

%only include the original form of the entities, and couple the method

%and stemming other terms using Porter stemmer, as described in ECS.1, it achieves similar performance to Porter stemmer and no statis

%For the outcomes of our proposed entity-based techniques, namely ECS.1 and ECS.2. First, we note a consistent trend across all ECS experiments: both the ChatGPT and LlaMa-2 based VS stemming methods demonstrate worse effectiveness compared to the Porter stemming approach. This reinforces the notion that LLM-based stemmers underperform the traditional Porter stemmer.

%Regarding the entity extraction models, Roberta displays comparable effectiveness to LlaMa-2 in ECS.1, but lags behind LlaMa-2 in ECS.2. Additionally, ECS.2 consistently outperforms ECS.1, underscoring the significance of incorporating entity tokens in both their original and stemmed forms. Notably, the only pipeline that exhibits statistically significant improvement over the Porter stemmer is ECS.2 when paired with the LlaMa-2 entity extraction model and the Porter stemmer. These findings strongly advocate for the potential utility of LLMs in the text pre-processing pipeline, particularly when employed as a entity extraction model.

%\include{gain_loss_analysis.tex}

\section{Conclusion}

We investigated whether using LLMs could improve the effectiveness of stemming when used within a keyword matching IR pipeline. With this respect, we devised three directions for using LLMs in a stemming pipeline.
%in IR, specifically focusing on their ability to enhance the effectiveness of lexical ranking methods like BM25. 
We found that \vs, although being the most computationally attractive method among the LLM-based alternatives, provided poor results compared to Porter stemming (though it did outperform Krovetz stemming). Similarly, also \cs, while intuitively promising, it provided poor effectiveness, at the expense of high computational costs. 

Our experimental results indicate that while \vs fails to outperform traditional stemming methods such as the Porter stemmer, its ability to extract entities that should not be stemmed could be used in conjunction with traditional stemming techniques to help with a more effective BM25 ranking. 
On the other hand, \cs, while theoretically intuitive, underperformed in our tests compared to traditional stemmers significantly. %This could be attributed to the inherent complexity of contextual stemming as a task. 
\ecs did provide significant improvements compared to using Porter alone: while incurring less computing costs than \cs (because of the lower number of output tokens being generated by the LLM), these are still considerable. %All code and results are available at \url{https://anonymous.4open.science/r/ECIR-2024-LLM-Stemming-847B}.

Our findings suggest that while LLMs may not currently be suitable for general stemming tasks within IR pipelines, at least when dealing with English text. While stemmers for English text have been well researched and developed, we note that stemmers for other languages often lack comparable attention, and effectiveness~\cite{silvello2018statistical} -- an interesting direction of future work is investigating the use of these foundational models as stemmers for languages other than English. The study has also opened up new avenues for future research: a promising direction is the integration of LLM-powered query term expansion within the ECS methods, potentially leading to further improvements in lexical retrieval effectiveness.

%These findings suggest that while LLMs may not currently be suitable for general stemming tasks in IR, they could serve a complementary role, enhancing the performance of existing stemming methods under specific conditions, such as extracting entities to avoid stemming. 

%Our findings have also opened up new avenues for future research. One promising direction could be the integration of query term expansion with the ECS method, potentially leading to further improvements in lexical retrieval effectiveness.

%\section*{Acknowledgment}
%Shuai Wang is supported by a UQ Earmarked PhD Scholarship and this research is funded by the Australian Research Council Discovery Projects programme ARC DP DP210104043.
\bibliographystyle{splncs04}
\bibliography{bibliography}

\begin{thebibliography}{10}
\providecommand{\url}[1]{\texttt{#1}}
\providecommand{\urlprefix}{URL }
\providecommand{\doi}[1]{https://doi.org/#1}

\bibitem{open-llm-leaderboard}
Beeching, E., Fourrier, C., Habib, N., Han, S., Lambert, N., Rajani, N.,
  Sanseviero, O., Tunstall, L., Wolf, T.: Open llm leaderboard.
  \url{https://huggingface.co/spaces/HuggingFaceH4/open_llm_leaderboard} (2023)

\bibitem{inpars}
Bonifacio, L., Abonizio, H., Fadaee, M., Nogueira, R.: {InPars}: Unsupervised
  dataset generation for information retrieval. In: Proceedings of the 45th
  International ACM SIGIR Conference on Research and Development in Information
  Retrieval. pp. 2387--2392. SIGIR '22, Association for Computing Machinery,
  New York, NY, USA (2022). \doi{10.1145/3477495.3531863},
  \url{https://doi.org/10.1145/3477495.3531863}

\bibitem{chang2023survey}
Chang, Y., Wang, X., Wang, J., Wu, Y., Zhu, K., Chen, H., Yang, L., Yi, X.,
  Wang, C., Wang, Y., et~al.: A survey on evaluation of large language models.
  arXiv preprint arXiv:2307.03109  (2023)

\bibitem{conneau2019unsupervised}
Conneau, A., Khandelwal, K., Goyal, N., Chaudhary, V., Wenzek, G., Guzm{\'a}n,
  F., Grave, E., Ott, M., Zettlemoyer, L., Stoyanov, V.: Unsupervised
  cross-lingual representation learning at scale. arXiv preprint
  arXiv:1911.02116  (2019)

\bibitem{croft2010search}
Croft, W.B., Metzler, D., Strohman, T.: Search engines: Information retrieval
  in practice, vol.~520. Addison-Wesley Reading (2010)

\bibitem{dai2019firstp}
Dai, Z., Callan, J.: Deeper text understanding for ir with contextual neural
  language modeling. In: Proceedings of the 42nd International ACM SIGIR
  Conference on Research and Development in Information Retrieval. pp.
  985--988. SIGIR'19, Association for Computing Machinery, New York, NY, USA
  (2019). \doi{10.1145/3331184.3331303},
  \url{https://doi.org/10.1145/3331184.3331303}

\bibitem{dai2023promptagator}
Dai, Z., Zhao, V.Y., Ma, J., Luan, Y., Ni, J., Lu, J., Bakalov, A., Guu, K.,
  Hall, K., Chang, M.W.: Promptagator: Few-shot dense retrieval from 8
  examples. In: The Eleventh International Conference on Learning
  Representations (2023), \url{https://openreview.net/forum?id=gmL46YMpu2J}

\bibitem{dietz2017component}
Dietz, F., Petras, V.: A component-level analysis of an academic search test
  collection. part i: system and collection configurations. In: Experimental IR
  Meets Multilinguality, Multimodality, and Interaction: 8th International
  Conference of the CLEF Association, CLEF 2017, Dublin, Ireland, September
  11--14, 2017, Proceedings 8. pp. 16--28. Springer (2017)

\bibitem{faggioli2023perspectives}
Faggioli, G., Dietz, L., Clarke, C.L., Demartini, G., Hagen, M., Hauff, C.,
  Kando, N., Kanoulas, E., Potthast, M., Stein, B., et~al.: Perspectives on
  large language models for relevance judgment. In: Proceedings of the 2023 ACM
  SIGIR International Conference on Theory of Information Retrieval. pp. 39--50
  (2023)

\bibitem{gao2022precise}
Gao, L., Ma, X., Lin, J., Callan, J.: Precise zero-shot dense retrieval without
  relevance labels. arXiv preprint arXiv:2212.10496  (2022)

\bibitem{inparsv2}
Jeronymo, V., Bonifacio, L., Abonizio, H., Fadaee, M., Lotufo, R., Zavrel, J.,
  Nogueira, R.: {InPars-v2}: Large language models as efficient dataset
  generators for information retrieval (2023). \doi{10.48550/ARXIV.2301.01820},
  \url{https://arxiv.org/abs/2301.01820}

\bibitem{kim2023solar}
Kim, D., Park, C., Kim, S., Lee, W., Song, W., Kim, Y., Kim, H., Kim, Y., Lee,
  H., Kim, J., Ahn, C., Yang, S., Lee, S., Park, H., Gim, G., Cha, M., Lee, H.,
  Kim, S.: Solar 10.7b: Scaling large language models with simple yet effective
  depth up-scaling (2023)

\bibitem{kopf2023openassistant}
K{\"o}pf, A., Kilcher, Y., von R{\"u}tte, D., Anagnostidis, S., Tam, Z.R.,
  Stevens, K., Barhoum, A., Duc, N.M., Stanley, O., Nagyfi, R., et~al.:
  Openassistant conversations--democratizing large language model alignment.
  arXiv preprint arXiv:2304.07327  (2023)

\bibitem{krovetz1993viewing}
Krovetz, R.: Viewing morphology as an inference process. In: Proceedings of the
  16th Annual International {{ACM SIGIR}} Conference on {{Research}} and
  Development in Information Retrieval. pp. 191--202. {ACM} (1993)

\bibitem{lin2023adapting}
Lin, D., Jing, L., Song, X., Liu, M., Sun, T., Nie, L.: Adapting generative
  pretrained language model for open-domain multimodal sentence summarization.
  In: Proceedings of the 46th International ACM SIGIR Conference on Research
  and Development in Information Retrieval. pp. 195--204 (2023)

\bibitem{lin2021pyserini}
Lin, J., Ma, X., Lin, S.C., Yang, J.H., Pradeep, R., Nogueira, R.: Pyserini: A
  python toolkit for reproducible information retrieval research with sparse
  and dense representations. In: Proceedings of the 44th International ACM
  SIGIR Conference on Research and Development in Information Retrieval. pp.
  2356--2362 (2021)

\bibitem{ma2023zero}
Ma, X., Zhang, X., Pradeep, R., Lin, J.: Zero-shot listwise document reranking
  with a large language model. arXiv preprint arXiv:2305.02156  (2023)

\bibitem{penedo2023refinedweb}
Penedo, G., Malartic, Q., Hesslow, D., Cojocaru, R., Cappelli, A., Alobeidli,
  H., Pannier, B., Almazrouei, E., Launay, J.: The refinedweb dataset for
  falcon llm: outperforming curated corpora with web data, and web data only.
  arXiv preprint arXiv:2306.01116  (2023)

\bibitem{peng2007context}
Peng, F., Ahmed, N., Li, X., Lu, Y.: Context sensitive stemming for web search.
  In: Proceedings of the 30th annual international ACM SIGIR conference on
  Research and development in information retrieval. pp. 639--646 (2007)

\bibitem{porter1980algorithm}
Porter, M.F.: An algorithm for suffix stripping. Programming  \textbf{14}(3)
  (1980)

\bibitem{porter2001snowball}
Porter, M.F.: Snowball: A language for stemming algorithms (2001)

\bibitem{pradeep2023rankvicuna}
Pradeep, R., Sharifymoghaddam, S., Lin, J.: Rankvicuna: Zero-shot listwise
  document reranking with open-source large language models. arXiv preprint
  arXiv:2309.15088  (2023)

\bibitem{qin2023large}
Qin, Z., Jagerman, R., Hui, K., Zhuang, H., Wu, J., Shen, J., Liu, T., Liu, J.,
  Metzler, D., Wang, X., et~al.: Large language models are effective text
  rankers with pairwise ranking prompting. arXiv preprint arXiv:2306.17563
  (2023)

\bibitem{sachan-etal-2022-improving}
Sachan, D., Lewis, M., Joshi, M., Aghajanyan, A., Yih, W.t., Pineau, J.,
  Zettlemoyer, L.: Improving passage retrieval with zero-shot question
  generation. In: Proceedings of the 2022 Conference on Empirical Methods in
  Natural Language Processing. pp. 3781--3797. Association for Computational
  Linguistics, Abu Dhabi, United Arab Emirates (Dec 2022).
  \doi{10.18653/v1/2022.emnlp-main.249},
  \url{https://aclanthology.org/2022.emnlp-main.249}

\bibitem{silvello2018statistical}
Silvello, G., Bucco, R., Busato, G., Fornari, G., Langeli, A., Purpura, A.,
  Rocco, G., Tezza, A., Agosti, M.: Statistical stemmers: A reproducibility
  study. In: Advances in Information Retrieval: 40th European Conference on IR
  Research, ECIR 2018, Grenoble, France, March 26-29, 2018, Proceedings 40. pp.
  385--397. Springer (2018)

\bibitem{singhal2023towards}
Singhal, K., Tu, T., Gottweis, J., Sayres, R., Wulczyn, E., Hou, L., Clark, K.,
  Pfohl, S., Cole-Lewis, H., Neal, D., et~al.: Towards expert-level medical
  question answering with large language models. arXiv preprint
  arXiv:2305.09617  (2023)

\bibitem{sun2023chatgpt}
Sun, W., Yan, L., Ma, X., Ren, P., Yin, D., Ren, Z.: Is chatgpt good at search?
  investigating large language models as re-ranking agent. arXiv preprint
  arXiv:2304.09542  (2023)

\bibitem{thomas2023large}
Thomas, P., Spielman, S., Craswell, N., Mitra, B.: Large language models can
  accurately predict searcher preferences. arXiv preprint arXiv:2309.10621
  (2023)

\bibitem{touvron2023llama}
Touvron, H., Lavril, T., Izacard, G., Martinet, X., Lachaux, M.A., Lacroix, T.,
  Rozi{\`e}re, B., Goyal, N., Hambro, E., Azhar, F., et~al.: Llama: Open and
  efficient foundation language models. arXiv preprint arXiv:2302.13971  (2023)

\bibitem{touvron2023llama2}
Touvron, H., Martin, L., Stone, K., Albert, P., Almahairi, A., Babaei, Y.,
  Bashlykov, N., Batra, S., Bhargava, P., Bhosale, S., et~al.: Llama 2: Open
  foundation and fine-tuned chat models. arXiv preprint arXiv:2307.09288
  (2023)

\bibitem{ws2023chatgpt}
Wang, S., Scells, H., Koopman, B., Zuccon, G.: Can chatgpt write a good boolean
  query for systematic review literature search? In: Proceedings of the 46th
  International ACM SIGIR Conference on Research and Development in Information
  Retrieval. pp. 1426--1436. SIGIR '23, Association for Computing Machinery,
  New York, NY, USA (2023). \doi{10.1145/3539618.3591703},
  \url{https://doi.org/10.1145/3539618.3591703}

\bibitem{wang2023generating}
Wang, S., Scells, H., Potthast, M., Koopman, B., Zuccon, G.: Generating natural
  language queries for more effective systematic review screening
  prioritisation. arXiv preprint arXiv:2309.05238  (2023)

\bibitem{wang2024zero}
Wang, S., Scells, H., Zhuang, S., Potthast, M., Koopman, B., Zuccon, G.:
  Zero-shot generative large language models for systematic review screening
  automation. arXiv preprint arXiv:2401.06320  (2024)

\bibitem{yang2023large}
Yang, C., Wang, X., Lu, Y., Liu, H., Le, Q.V., Zhou, D., Chen, X.: Large
  language models as optimizers. arXiv preprint arXiv:2309.03409  (2023)

\bibitem{yang2023exploring}
Yang, X., Li, Y., Zhang, X., Chen, H., Cheng, W.: Exploring the limits of
  chatgpt for query or aspect-based text summarization. arXiv preprint
  arXiv:2302.08081  (2023)

\bibitem{zhuang2023augmenting}
Zhuang, S., Shou, L., Zuccon, G.: Augmenting passage representations with query
  generation for enhanced cross-lingual dense retrieval. arXiv preprint
  arXiv:2305.03950  (2023)

\bibitem{zhuang2023setwise}
Zhuang, S., Zhuang, H., Koopman, B., Zuccon, G.: A setwise approach for
  effective and highly efficient zero-shot ranking with large language models.
  arXiv preprint arXiv:2310.09497  (2023)

\end{thebibliography}

%\section*{Appendix}

\end{document}